# A Hierarchical Key Management Scheme for Wireless Sensor Networks Based on Identity-based Encryption

Hu Shuaiqi
School of Information Science and Engineering
Southeast University
Nanjing, China
shuaiqihu@yahoo.com

*Abstract*—Limited resources (such as energy, computing power, storage, and so on) make it impractical for wireless sensor networks (WSNs) to deploy traditional security schemes. In this paper, a hierarchical key management scheme is proposed on the basis of identity-based encryption (IBE).This proposed scheme not only converts the distributed flat architecture of the WSNs to a hierarchical architecture for better network management but also ensures the independence and security of the sub-networks. This paper firstly reviews the identity-based encryption, particularly, the Boneh-Franklin algorithm. Then a novel hierarchical key management scheme based on the basic Boneh-Franklin and Diffie-Hellman (DH) algorithms is proposed. At last, the security and efficiency of our scheme is discussed by comparing with other identity-based schemes for flat architecture of WSNs.

*Keywords- WSNs; key management; IBE; Diffie-Hellman key exchange*

## I. INTRODUCTION

As one of the key technology in Internet of things (loT), wireless sensor networks has been widely used in military reconnoiter, city management, as well as environment and traffic detection. However, since sensor nodes and communication links in WSNs is open and unguarded, any potential adversary is likely to eavesdrop or fabricate the information being transmitted. Therefore, how to ensure network security and information authenticity is a technique that directly influences whether WSNs can be widely applied.

Traditionally, WSNs have two kind of network architecture: the distributed flat architecture and the hierarchical architecture [1]. In a hierarchical architecture of WSNs, nodes can be divided into two kinds: the cluster heads and sensor nodes. Cluster heads are superior to sensor nodes in terms of computational ability, storage, and battery power. In a distributed flat architecture of WSNs, every nodes is entitled the same status with same resources. However, because the WSNs that are deployed hierarchical architecture must ensure the designated cluster heads to be in the mist of the networks to avoid unattended nodes, the application of hierarchical architecture WSNs is greatly limited in practice. In many application areas such as the battlefield, harsh natural environment, distributed flat architecture is widely used in WSNs.

In order to ensure the security of distributed flat architectural WSNs, researchers have been investigating a variety of security schemes. All these security schemes can be divided into two types: symmetric cryptography or asymmetric cryptography.

Compared to asymmetric cryptography, symmetric cryptography requires far less computation. Therefore, symmetric cryptography has attracted much attentions and many schemes has been proposed. Eschenauer and Gligor has proposed a scheme based on probability for pre-distribution of keys [2]. Based on their scheme, Pietro develops a random key distribution scheme [3].

However, in order for symmetry cryptography to be deployed in WSNs, a pre-distribution process of keys must be finished and each node must store symmetry keys before the entire network is deployed, which would cause a big trouble for nodes addition or revocation. Thereby, researchers has been trying to deploy asymmetry cryptography in WSNs. Identity-based Encryption (IBE) [4], proposed in 1984 by Shamir, has been attached great importance because of its advantages in discernable pubic keys, not requiring PKI to distribute certificates and low computation overhead.

Yang proposed a session key agreement scheme suitable for WSNs based on IBE and Diffie-Hellman key exchange [5]. In his scheme, key distribution and negotiation only happen in the initialization process of the networks, and once session keys have been established, information will be encrypted by symmetry cryptography, which requires less computation overhead. Due to its simplicity, this scheme is applicable to distributed flat architecture WSNs. Inspired by this scheme, Guo makes some improvement on Yang's scheme in terms of energy consumption and computation overhead by using the identity-based signature to replace the identity-based encryption [6]. Nevertheless, both scheme fails to realize the authentication of nodes, and because each nodes in the network has to negotiate session keys with all neighborhood nodes, the communication and computation overhead is considerable. Qin designed a scheme for hierarchical architecture WSNs [7], which entitles sensor nodes to play the role of cluster heads, but this scheme is only aimed for hierarchical architectural WSNs and the computation overhead is intolerable for ordinary nodes since every communication process requires the encryption by IBE and symmetric cryptography.

In this paper, we propose a hierarchical key management scheme based on IBE and DH key exchange. This scheme can greatly decrease the frequency of key negotiation and thus decrease the communication overhead by converting the distributed flat architecture to hierarchical architecture. The information will be encrypted by symmetry encryption rather than complex IBE, and hence subtly avoid the computation overhead brought by IBE.

The rest of the article is organized as follows. Section II describes the fundamental properties of identity-based encryption and its mathematical basis. Section III proposes a hierarchical key management scheme based on IBE and DH key exchange. Section IV & V give a detailed analysis of the scheme in terms of security and performance. The paper ends in section V with some conclusions and the expectation for future works.

## II. IDENTITY-BASED ENCRYPTION

Identity-based encryption forsakes obtaining public key from certificate of the PKI, which is used in traditional asymmetric cryptography algorithm. In IBE, a string related to user's identity is used as the public key of the user. Considering the computation overhead and security, this paper selected the identity-based encryption algorithm proposed by Boneh and Franklin [8]. The following parts of this section would introduce this IBE algorithm and relevant mathematic background.

### A. Secutity Model

The security of the algorithm proposed in [8] lies on the Bilinear Diffie-Hellman (BDH) problems. The core of this algorithm is to create a Weil pairing in supersingular elliptic curve. The BDH problem and Weil pairing will be described in the following.

#### 1) BDH problem

For random numbers a, b, c $\in Z_P^*$ , compute $\hat{e}(P,P)^{abc}(GF(p^2))$ on the assumption that (a, aP, bP, cP) is known, and $\hat{e}: G \times G \to GF(p^2)$ is a mapping that has the following proprieties:

  *a) Bilinear.* If for all $x, y \in G$, $a, b \in Z$, there will be $\hat{e}(x^a, y^b) = \hat{e}(x,y)^{ab}$, we can regard the map $\hat{e}$ as bilinear.

  *b) Nondegenerate.* There exists a pair of P, Q$\in G$ to make $\hat{e}(P,Q) \neq 1$.

  *c) Computable.* For any P, Q$\in G$, there is a effective algorithm to compute $\hat{e}(P,Q)$.

#### 2) Weil pairing

The definition of Weil pairing is described as following: assume the order of cyclic group G is q, and there is an equation $\gcd(m, q) = 1$, then the Weil pairing with order m is a mapping that fulfills following requirements:

$$e_m: \begin{cases} E[m] \times E[m] \to F_q^{k*} \\ (P,Q) \to f_A(B)/f_B(A) \end{cases},$$

In this mapping, P，Q$\in E[m]$, A ~ (P) ~ (O), B ~ (Q) – (O), $(f_A) = mA$, $(f_B) = mB$.

### B. Basic Boneh-Franklin IBE algorithm

The Basic Boneh-Franklin IBE algorithm is comprised of four subsystems: setup, extract, encrypt and decrypt. These four subsystems will be introduced on the assumption that plaintext space M = $\{0,1\}^n$, cipher space C = $G_1 \times \{0,1\}^n$.

  *1) Setup.*

  *a)* Private Key Generator (PKG) chooses a supersingular elliptic curve that fulfills WDH security assumption and a large prime p that is k bits long. Then PKG generates a subgroup G with order q in $E/GF(p)$, G's generator P, as well as a bilinear mapping $\hat{e}: G \times G \to GF(p^2)$.

  *b)* PKG picks a random master key $s \in Z_q^*$ , and computes $P_{pub} = sP$.

  *c)* Select two hash function $H_1: \{0,1\}^* \to E/GF(P)$, $H_2: GF(p^2) \to \{0,1\}^n$ . Construct the output public parameter $\pi = \{q, p, \hat{e}, n, P, P_{sub}, H_1, H_2\}$, master key $s$ is kept only by PKG.

  *2) Extract*

  For a given string Id $\in \{0,1\}^*$, compute $Q_{Id} = H_1(Id) \in E/GF(p)$, and compute $K_{Id} = (Q_{Id})^s$ as private key for corresponding user.

  *3) Encryption*

  For a plaintext m $\in$ M, and its Id, the method to encrypt it is to: first, compute $Q_{Id} = H_1(Id) \in GF(p)$; second, choose a random number r $\in Z_q^*$; third, construct ciphered text C =< rP, m $\oplus H_2(g_{Id}^r) >$ , among which $g_{Id} = \hat{e}(Q_{Id}, P_{pub}) \in GF(p^2)$.

  *4) Decryption*

  Assuming that the ciphered text C =< U, V >, apply the private key $K_{Id}(E/GF(p))$ to compute the plaintext m = V $\oplus H_2(\hat{e}(K_{Id}, U))$.

## III. A HIERARCHICAL KEY MANAGEMENT SCHEME FOR DISTRIBUTED FLAT ARCHITECTURAL WSNS

Based on Basic Boneh Franklin algorithm, we propose a hierarchical key management scheme. The core of this scheme is to manage the distributed flat architecture WSNs in a hierarchical way. The major technique used in this scheme is to allow some sensor nodes to serve as cluster heads routinely and dynamically. Other sensor nodes negotiate symmetry keys with cluster heads by applying IBE algorithm and Diffie-Hellman key exchange. In addition, nodes no longer use IBE algorithm to encrypt message since the computation overhead is intolerable for nodes with limited recourses, instead, they use symmetric cryptography to encrypt the message to be sent. The following parts of this section introduce our scheme in terms of network model, network initialization, key negotiation, node addition and revocation.

### A. Network model

As said before, this scheme is aimed for distributed flat architectural WSNs, and achieves a hierarchical management through the dynamic allocation of cluster heads. By default, a base station is in the center of the network and all other nodes are randomly distributed and accessible by the base station. Considering the geographical distribution of nodes, the base station chooses N sensor nodes to be cluster heads, which forms N sub-networks. The cluster heads chosen by base station delimit their territories and broadcast their identities to sensor nodes within their territories. The sensor node in a specific territory communicate only with its cluster head. The cluster heads transmit information received from sensor nodes to base station and make a rudimentary information fusion. If the base station is accessible to cluster heads, cluster heads transmit messages directly to base station, if not, cluster heads make a multi-hop transmission via reply nodes. The network architecture is illustrated in Figure 1.

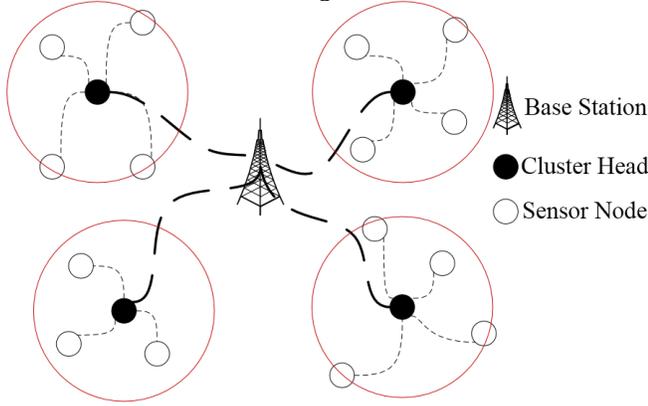

Figure 1. A hierarchical architecture for distribute flat WSNs.

### B. Network initialization

During the manufacture period of sensor nodes, the *setup* function of IBE is executed and all sensor nodes store the public parameter π. Master key *s* is only accessible to the owner of the network or the base station. For example, if a troop wants to deploy a WSN in a battlefield, the public parameter π should be embedded in the ROM of every nodes during the manufacture period. However, the master key should be preserved by the troop or the base station owned by this troop. If new sensor nodes need to be added to the network, we must make sure that the parameters embedded in the new sensor nodes are identical to that of the deployed nodes. In addition, before nodes are put into use, a unique Id number should be allocated to every nodes and the private key $K_{Id}$ corresponding to Id number should be generated and allocated to nodes by calling the *extract* function of IBE.

After the network is deployed, the base station chooses N sensor nodes to be cluster heads and thus N sub-networks is formed. Once the nodes chosen know about their identity as cluster heads, they negotiate secret keys with base station. Meanwhile, cluster heads broadcast their Ids to neighborhood sensor nodes and negotiate secret keys with sensor nodes within their territories. The detailed process of key negotiation between cluster heads and base station or sensor nodes is discussed in part C, section III.

After the negotiation of keys between sensor nodes and cluster heads is finished, a cluster head knows the Id of every sensor nodes in its territory and acquires communication keys with every associated nodes. Then a cluster head choose a random number $s_i \in Z_q^*(i = 1,2,3,…,N)$ as a master key and serve as PKG in its territory. The cluster head uses the master key $s_i$ and each node's Id to generate new private keys for every nodes in its territory, then $P_{pub} = sP$ will be calculated. The newly calculated private key $K_{Id}$ and $P_{pub}$ are updated in cluster heads. Besides, each node would receive encrypted $P_{pub}$ and $K_{Id}$ that are corresponding to their Ids from cluster heads. After receiving the message, each node decrypts it and update their private key and public parameter.

Therefore, the entire network is divided into N separate and independent sub-networks through the function of cluster heads. In each sub-network, a cluster head serves as PKG and can carry on the key distribution and negotiation process independently. Since the master key $s_i$ is randomly chosen by each cluster head, every two sub-networks are irrelevant. Granted that the adversary has made a breakthrough in one sub-network, other sub-networks are still secure.

### C. Key negotiation

The key negotiation process consists of three steps: broadcast, parameter calculation, parameter exchange. The calculation of symmetry keys is proceeded using Diffie-Hellman key exchange algorithm [9]. The entire negotiation process is illustrated in figure 2.

*1) Broadcast*

Once a cluster head (denoted as A) knows its identity, it broadcasts its identification $Id_A$ to base station and sensor nodes in its territory (denoted as B).

*2) Parameter calculation*

A cluster head (A) chooses $X_A < q$, and calculates parameter $Y_A = \eta^{X_A} \bmod q$.

A base station or sensor node (B) chooses $X_B < q$, and calculates parameter $Y_B = \eta^{X_B} \bmod q$.

*3) Parameter exchange*

*a)* A base station or sensor node uses the *encrypt* function of IBE algorithm to encrypt $Y_B$ and $Id_B$. The public key used for encryption is the identity of the cluster head: $Id_A$, which is broadcasted by the cluster head.

*b)* Once the cluster head receives the encrypted message, it uses its private key $K_{Id_A}$ to *decrypt* the message and get $Y_B$ and $Id_B$ from the decrypted message.

*c)* Correspondingly, the cluster head uses the *encrypt* function of IBE algorithm to encrypt $Y_A$. The public key used for encryption is the identity of the node or base station: $Id_B$. Then the cluster head sends the encrpypted message to the corresponding receiver, that is the node or base station which owns the private key of the corresponding public key.

*d)* The cluster head calculates and stores the symmetric key $K = (Y_B)^{X_A} \bmod q$, the base station or sensor node calculates and stores the key $K = (Y_A)^{X_B} \bmod q$, too.

*4) Encryption and Decryption process*

Communication parties use the symmetry key K as the secret key to encrypt or decrypt message, using symmetry cryptography algorithm such as AES or DES. After a certain period of time, in order to ensure the security of the network, each sub-network proceeds independent key distribution and negotiation process. During this process, the public parameter and private keys used for key distribution or negotiation is the updated value, which could make sure the independence of each sub-network.

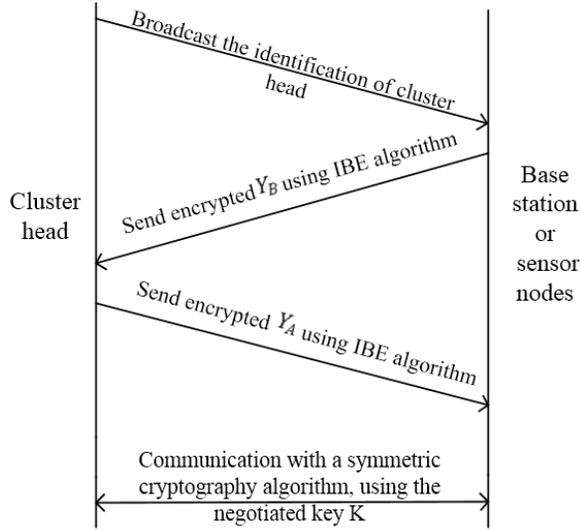

Figure 2. Key negotiation process

*D. Node addition and revocation*

Since the public parameter π has already been stored in the ROM of each node, new nodes are entitled secret keys to communicate with cluster heads. If a new node needs to be added to the network, it ought to register his identity to cluster head in the sub-network, which it belongs to. Then the cluster head checks the validity and authenticity of the new node through communicating with the base station. If this node is valid and reliable, cluster head would send the updated public parameter and the new node's new private key to it, using the IBE. Once the new node has received and updated the private key and public parameter, it negotiate with the cluster head to form a communication key. The detail of the key negotiation is illustrated in part C, section III.

If a node needs to be annulled, the cluster head discards the corresponding communication key and reports the identity of the annulled node to base station.

IV. SECURITY ANALYSIS

In this section, the security of our scheme is discussed in terms of compromised nodes, the security of the parameter exchange protocol and forward security.

*A. Compromised nodes*

Compromised nodes are nodes that are manipulated by the adversary through some kind of techniques. A secure WSNs scheme should make it difficult or impossible for adversary to extract information about other nodes or the network through compromised nodes. In our scheme, since every node has negotiated a unique and independent key with the cluster head, even though the adversary captures some sensor nodes, the keys between the cluster head and other nodes are still secure. Granted that a cluster head is captured by the adversary, the base station can detect the behavior of cluster heads and thus find out the comprised nodes. Once a cluster head found compromised, the base station designates another node to serve as a cluster head. The new cluster head regenerates and redistributes sub-network's public parameter and each node's private key. Since the new cluster head can refuse to communicate with the compromised former cluster head, the compromised cluster head would be ruled out of the network and the adversary can get no further information about the network. Furthermore, the adversary is not able to get the communication keys between cluster head and sensor nodes through analyzing the ciphered text for its difficulty is identical to solve the Discrete Logarithm Problem (DLP). In addition, acquiring the private keys of nodes is also regarded as impossible, since its difficulty is identical to solve the Bilinear Diffie-Hellman (BDH) Problem.

*B. The security of the parameter exchange protocol*

In our scheme, a secrete key is not decided by a single party. Instead, it is negotiated through Diffie-Hellman key exchange with several advantages of the key agreement protocol [11]. In addition, since the exchange parameter of Diffie-Hellman key exchange is encrypted by IBE algorithm through the receiver's public key, only the designated user can decrypt the message and thus attain an implicit authentication. Furthermore, man-in-the-middle attack is not practical in our scheme since every parameter is encrypted and can only be decrypted by the receiver.

*C. Forward Security*

Forward security means the long time leakage of keys from one or more entities could not lead to the leakage of the session keys. In our scheme, the cluster head is dynamically assigned by the base station rather than changeless. The management of each sub-network is also independent. Even though the private keys of a sub-network is leaked, through the dynamic adjustments of the cluster head, the public parameter of the sub-network would be adjusted and thus the communication keys between the cluster head and sensor nodes would be adjusted. Therefore, even though the private keys are leaked, the adversary can only get information in a certain region during a limited time span. The information transmitted before the former round of key distribution as well as after a new round of key distribution is secure. In conclusion, our scheme can achieve approximate perfect forward security.

V. PERFORMANCE ANALYSIS

In this section, we will discuss the performance of our scheme and compare the computation overhead and communication overhead with IBEKAS scheme and BNN-IBS-KS scheme, which is illustrated respectively in [8] and [9]. The following analysis and comparison is based on these assumptions: 1) there are N sub-networks in a network, 2) each sub-network contains M nodes, 3) there are also M nodes in the communication range of each node.

## A. Connectivity rate

According to our scheme, the key agreement between the cluster head and sensor nodes or base station is implemented through the directional transmission of the encrypted DH key exchange parameter, which is encrypted by the receiver's public key. Base station or each nodes can negotiate a symmetry key with the cluster head. In addition, the base station can cluster the network to make sure the cluster heads can cover all the nodes, by analyzing each node's geographical location and communication capacity. Thereby, theoretically, our scheme can achieve approximately 100% connectivity rate. However, it is very difficult for key pre-distribution scheme to achieve high connectivity rate due to the randomness of the key distribution or the haphazard deployment of nodes [6].

## B. Network Flexibility

The process of node addition and revocation has been discussed in part D, section III. According to IBEKAS scheme or BNN-IBS-KS scheme, a new node must conduct key agreements between all neighborhood nodes while being added to the network. However, in our scheme, only the new node itself, the cluster head that rules the region new node belongs to and the base station are involved in the process of node addition. Other nodes are free from the intervention of new node addition and thus the network load is decreased.

If a certain node needs to be revoked from the network, the base station would preserve its identification. Since the cluster head must register the new node's identification to the base station before this node is able to join the network, nodes that have been revoked is impossible to join the network again, which forbids the adversary to use the revoked nodes to imitate new nodes. However, there is no mechanism in IBEKAS scheme or BNN-IBS-KS scheme to prevent the adversary from imitating new nodes with revoked nodes, which makes it pretty easy for the adversary to use the revoked nodes to endanger network security.

## C. Comunication overhead

In our scheme, every cluster head negotiates secret keys with base station and each node in its territory. The total negotiation times in a sub-network is M+1. However, in IBEKAS scheme and BNN-IBS-KS scheme, each node must negotiate secrete keys with every nodes in its neighborhood, and thus $C_M^2 = (M \times (M-1))/2$ times key agreements must be conducted. The comparison between our scheme and IBEKAS, BNN-IBS-KS is illustrated in figure 3. In WSNs, energy consumed by wireless transmission is far more than the energy consumed by computation [10]. Thus compared to IBEKAS and BNN-IBS-KS, our scheme can greatly save the energy consumed by wireless transmission.

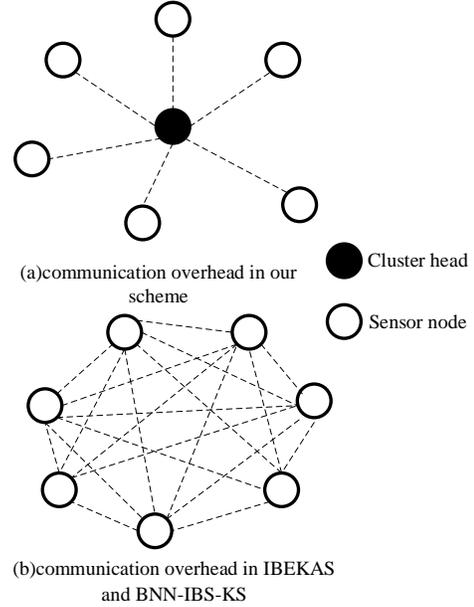

(a) communication overhead in our scheme

(b) communication overhead in IBEKAS and BNN-IBS-KS

Figure 3. Comparison of communication overhead

## D. Computation overhead

In each sub-network, our scheme requires M+1 times key agreement processes. However, IBEKAS and BNN-IBS-KS requires $(M \times (M-1))/2$ times key agreement processes. The computation overhead of our scheme, IBEKAS and BNN-IBS-KS in a sub-network that contains M nodes is described in Table I.

TABLE I. COMPUTATION OVERHEAD COMPARISION

| Arithmetic Type | Proposed scheme | | IBEKAS | | BNN-IBS-KS | |
|---|---|---|---|---|---|---|
| | *Encryption* | *Decryption* | *Encryption* | *Decryption* | *Signature* | *Validation* |
| Bilinear | M+1 | M+1 | $\frac{M(M-1)}{2}$ | $\frac{M(M-1)}{2}$ | | |
| Hash | 2M+2 | M+1 | $M(M-1)$ | $\frac{M(M-1)}{2}$ | $\frac{M(M-1)}{2}$ | $M(M-1)$ |
| ECC addition | M+1 | | $\frac{M(M-1)}{2}$ | | $\frac{M(M-1)}{2}$ | $\frac{3M(M-1)}{2}$ |
| ECC multiplication | | | | | | $M(M-1)$ |
| XOR | M+1 | M+1 | $\frac{M(M-1)}{2}$ | $\frac{M(M-1)}{2}$ | | |
| Exponent | M+1 | | $\frac{M(M-1)}{2}$ | | | |

The computation overhead in IBE is mainly from bilinear operation and ECC multiplication operation. To simplify analysis, we only consider these two operations. In a MICA2 sensor (8bits, ATmega128L, 8MHz, voltage 3V, current 8mAh), implementing an ECC multiplication operation requires 0.81s and 19.44mJ energy. Implementing a bilinear operation requires 3.102s and 74.45mJ energy [6]. In a sub-network with M nodes, the computation time and energy consumption is illustrated in table 2.

TABLE II. COMPUTATION TIME AND ENERGY CONSUMPTION

| Scheme | Time (s) | Energy (mJ) |
|---|---|---|
| Proposed scheme | $(0.81 + 3.102 \times 2) \times M$ | $(19.44 + 74.45 \times 2) \times M$ |
| IBEKAS | $(0.81 + 3.102 \times 2) \times \frac{M(M-1)}{2}$ | $(19.44 + 74.45 \times 2) \times \frac{M(M-1)}{2}$ |
| BNN-IBS-KS | $0.81 \times (\frac{3M(M-1)}{2} + 1)$ | $19.44 \times (\frac{3M(M-1)}{2} + 1)$ |

With different nodes in a sub-network, the computation time and energy of the three schemes is presented in figure 4.

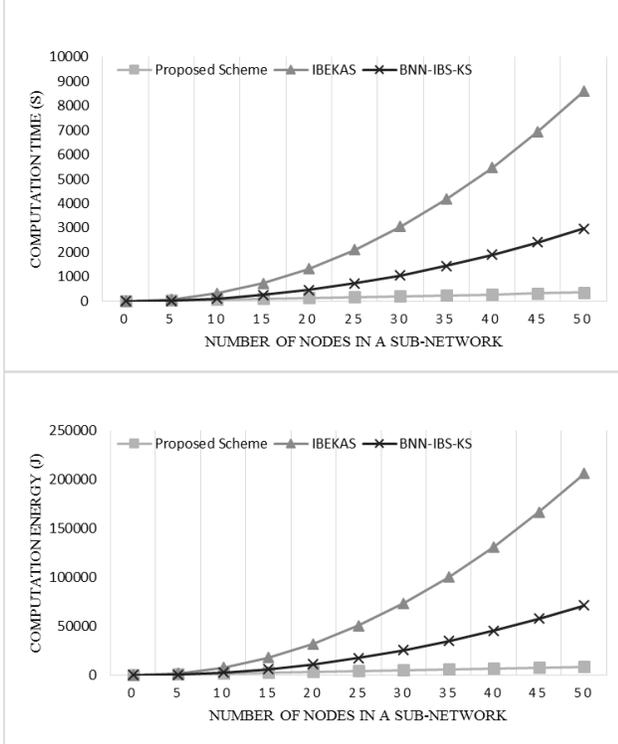

Figure 4. Computation time and energy consumption

According to the comparison in figure 4. It is clear that, compared to IBEKAS or BNN-IBS-KS, our scheme dramatically decreases the computation time and energy consumption through the decrease of negotiation times, which is critical for WSNs with limited computation resources and energy.

VI. CONCLUSION

This paper proposes a scheme to manage distributed flat WSNs in a hierarchical way. This scheme not only solves the large consumption in communication and computation confronted by traditional distributed flat WSNs, but also serves as an effective method to prevent the adversary from using the compromised nodes to threaten the entire network.

ACKNOWLEDGMENT

This work is supported by Southeast University Students Research Training Program (SRTP) under grant T14042007 and Information Security Center of Southeast University.